# Contraction of polymer gels created by the activity of molecular motors

Mattia Bacca[1*], Omar A. Saleh[2,3], Robert M. McMeeking[2,4,5,6]

[1]Mechanical Engineering Department, University of British Columbia, Vancouver, BC V6T 1Z4 – Canada
[2]Materials Department, University of California, Santa Barbara, CA 93106 – USA
[3]Biomolecular Science & Engineering Program, University of California, Santa Barbara, CA 93106 – USA
[4]Mechanical Engineering Departments, University of California, Santa Barbara, CA 93106 – USA
[5]School of Engineering, University of Aberdeen, King's College, Aberdeen AB24 3FX – UK
[6]INM—Leibniz Institute for New Materials, Campus D2 2, Saarbrücken, 66123 – Germany

*Abstract*
We propose a theory based on non-equilibrium thermodynamics to describe the mechanical behavior of an active polymer gel created by the inclusion of molecular motors in its solvent. When activated, these motors attach to the chains of the polymer network and shorten them creating a global contraction of the gel, which mimics the active behavior of a cytoskeleton. The power generated by these motors is obtained by ATP hydrolysis reaction, which transduces chemical energy into mechanical work. The latter is described by an increment of strain energy in the gel due to an increased stiffness. This effect is described with an increment of the cross-link density in the polymer network, which reduces its entropy. The theory then considers polymer network swelling and species diffusion to describe the transient passive behavior of the gel. We finally formulate the problem of uniaxial contraction of a slab of gel and compare the results with experiments, showing good agreement.

*Introduction*
Active polymer gels have been prototyped in the attempt to synthetically reproduce the active mechanical behavior of the cytoskeleton.  This is useful for mimicking cellular activities that lack genetic control and to prototype a new generation of active materials for a wide set of technological applications [1].  These gels are synthesized by polymerization of long hydrophilic chains forming a loose network that is capable of large swelling due to absorption of an aqueous solvent.  The solvent includes special proteins, *molecular motors*, which attach to the polymer chains. These proteins act as enzymes for ATP (adenosine triphosphate) hydrolysis reaction taking place within the solvent.  The reaction converts ATP molecules into ADP transducing chemical energy into mechanical work.  This mechanical work has the effect of shortening the polymer chains where the motors are attached.  Figure 1 provides a sketch showing the activity of a molecular motor shortening a polymer chain, with resulting reduction of the average spacing among crosslinks.  As a consequence, the gel becomes stiffer and solvent molecules will diffuse away to accommodate contraction, a mechanism that has also been observed in the cytoskeleton [2] and reproduced in gels made from in vitro cytoskeletal components [1, 3-5].

Kruse and coworkers [6] developed the first continuum theory (to the authors' knowledge) to describe the activity of molecular motors in the context of active polar gels. Their aim was to describe the behavior of a cytoskeleton. The effect of the motors, pulling on the polymer chains, is represented by force dipoles. These dipoles create a macroscopic effect that is described with a reactive stress, which is then added to the passive viscoelastic stress of the gel. The total stress then equilibrates all the external loads. More recently, MacKintosh and Levine [7] proposed a similar hydrodynamic theory based again on the generation of force dipoles by motor activity within the gel, which translates into the generation of a transient contractile stress.

The main limitation of these theories lies in the description of the macroscopic effect of the motor activity with a stress. Imagine a motor pulling on a very loose polymer chain as observed in experiments [1]. The macroscopic stress generated by this activity is negligible, yet the motor is consuming a meaningful amount of fuel (ATP molecules). The gel stiffness on the other hand is significantly affected. The transduction of chemical energy into mechanical work is thus internal to the material, and the polymer stiffens. This effect cause macroscopic contraction when solvent diffusion is allowed. We adopt the concept that the stiffness increment is due to evolution of the crosslink density since the average distance among crosslinks is reduced by the motors shortening the chains, as sketched in Figure 1. We take this point of view because it is likely that the polymer network in the gel is far from being in a stretched condition even after chain shortening by molecular motor activity. Therefore, network elasticity is probably entropic rather than enthalpic, and the work done by molecular motors reduces network entropy, increasing the system free energy. The additional stored energy in the system is associated with increased constraints on chain fluctuations, thereby stiffening the gel. Crosslink density evolution is created by a motor attaching to a single chain and shortening it, as described above, and it is also created by a motor attaching to two distinct chains, hence acting as a dynamic crosslink. Both effects can be described with the proposed model, however we will focus on the chain shortening mechanism. We describe the passive behavior of the gel by Flory's and Rehner's [8] theory for polymer network swelling, include species diffusion as proposed by Hong *et al.* [9], and add to that model an increase in crosslink density to account for the effect of molecular motors.

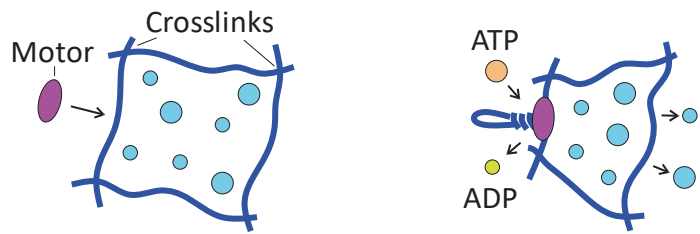

**Figure 1** Left: Molecular motor binding to a polymer chain (blue spline); Right: Motor shortening the polymer chains by conversion of ATP into ADP via hydrolysis and resulting in gel contraction by local expulsion of solvent molecules (cyan circles).

As observed by Bertrand *et al.* [1], some motors attach to the chains and keep them at a constant shortening ratio. These *static motors* are in their "reeled-in" state and give steady state stiffening to the gel, as much as a 10-fold increase in elastic modulus [1]. The remaining motors that attach to the polymer chains, and that are not in their reeled-in state, give active contraction by progressive incremental chain shortening. These *dynamic motors* attach to the chains, reel them in, but then detach, with many of them doing so simultaneously and continuously. The polymer chain being shortened by a motor exerts resistive forces on the latter because its fluctuations are being constrained. This effect challenges the strength of the bonds between motor and chain, and for some of the motors causes the bonds to be broken. This phenomenon explains why some motors detach from the chains after a period of reeling them in. As a consequence, the polymer chain extends back to its original length and the motor is free to move in the solvent until it bonds to another chain and starts the shortening process again. When a given motor detaches from the chain, its effect on the stiffness is lost and so is its contribution to the contraction of the gel. For simplicity we assume that all the motors that attach to and detach from the chains do so simultaneously and in phase so that there is periodic *contraction* and *recovery* in the gel. This hypothesis describes the behavior of local elements of the gel, and relies on small spatial variation of chain length and chain shortening rate.

*Thermodynamic framework*
We consider the reference state of the gel, of volume $V_0$ with surface $S_0$, to be that of the unswollen polymer network alone. The conditions are quasi-static, and thus the 1st Piola-Kirchhoff stress, $t_{ij}$, obeys

$$\frac{\partial t_{ij}}{\partial X_j} + B_i = 0 \qquad \text{in } V_o \tag{1a}$$

and

$$t_{ij} N_j = T_i \qquad \text{on } S_o \tag{1b}$$

where $N_i$ is the outward unit normal to $S_o$, $T_i$ is the surface traction, $X_i$ is the position of elements of the polymer network in the reference configuration, and $B_i$ is the body force per unit reference volume.

During deformation of the gel, including swelling, the current position of elements of the polymer network is given by

$$x_i = x_i(X_j, t) \tag{2}$$

where $t$ is time. The deformation gradient is then given by

$$F_{ij} = \frac{\partial x_i}{\partial X_j} \tag{3}$$

The solvent consists of water and other mobile species. Some of these species are reactive, some other species are inert, *i.e.* they do not participate in any chemical reaction. Let species $k$ have concentration $C^k$ in moles per unit volume of the reference state. Take the molar volume of species $k$ to be $\Omega^k$ so that the volume of fluid associated with $dV_o$ is $\Omega^k C^k dV_o$ (unless otherwise specified, a repeated index within a product indicates a sum). Molecular incompressibility of all the species involved imposes the constraint

$$J = 1 + \Omega^k C^k \tag{4}$$

where $J$ is the determinant of the deformation gradient $F_{ij}$, and is equal to the ratio $dV/dV_o$, with $dV$ the infinitesimal volume of $dV_o$ in the deformed configuration.

Conservation of species *k* requires that

$$\frac{dC^k}{dt} = Q_c^k - \frac{\partial J_i^k}{\partial X_i} \tag{5}$$

where $Q_c^k$ is a source for species *k* due to chemical reactions, measured in moles per unit reference volume, and $J_i^k$ is the flux of species *k*, identified in the reference configuration.

The Helmholtz energy per unit reference volume of the gel is assumed to be the functional

$$\psi = \psi(F_{ij}, T, C^k, N) \tag{6}$$

where $T$ is temperature, and $N$ is the cross-link density of the polymer network. As explained in *Appendix A*, thermodynamics dictates

$$t_{ij} = \frac{\partial \psi}{\partial F_{ij}} \tag{7}$$

$$\mu^k = \frac{\partial \psi}{\partial C^k} \tag{8}$$

and

$$\eta = -\frac{\partial \psi}{\partial T} \tag{9}$$

with $\mu^k$ the chemical potential of species *k* and $\eta$ the entropy per unit reference volume of the gel. We assume positive entropy production in the material to satisfy the second law of thermodynamics. Also, we assume that there are 3 independent processes, namely (i) chemical reactions and evolution of cross-link density, (ii) heat diffusion and (iii) mass diffusion. Positive entropy production, according to *Appendix A*, then gives the following 3 inequalities

$$\left(\frac{\partial \mu^k}{\partial X_i} + \eta^k \frac{\partial T}{\partial X_i}\right) J_i^k \leq 0 \tag{10}$$

which controls the direction of mass flux, with $\eta^k$ the entropy per mole of species *k*,

$$\frac{J_i^h}{T} \frac{\partial T}{\partial X_i} \leq 0 \tag{11}$$

which controls the direction of the heat flux $J_i^h$, and

$$\mu^k Q_c^k + \frac{\partial \psi}{\partial N} \frac{dN}{dt} \leq 0 \tag{12}$$

which controls the direction of chemical reactions in relation to an evolution of the cross-link density.

The chemical reaction considered is ATP hydrolysis, and thus Eq. (12) becomes

$$\frac{\partial \psi}{\partial N} \frac{dN}{dt} \leq \Delta G^h Q_c \tag{13a}$$

with

$$\Delta G^h = \mu^{\text{ATP}} + \mu^{\text{H}_2\text{O}} - \mu^{\text{ADP}} - \mu^{\text{P}} \tag{13b}$$

the Gibbs energy released by one mole of ATP ($32-40 \ kJ/mol$, [10]) undergoing the hydrolysis reaction

$$\text{ATP} + \text{H}_2\text{O} \rightarrow \text{ADP} + \text{P}$$

where P is the ATP lost phosphate. In Eq. (13a) $Q_c$ is the reaction rate, in moles of ATP consumed per second per unit reference volume. Considering $\varphi$ to be the average efficiency of all the motors in $V_0$ in transducing chemical energy into mechanical work [11], we deduce that

$$\frac{\partial \psi}{\partial N} \frac{dN}{dt} = \varphi \Delta G^h Q_c \tag{14}$$

where Eq. (13a) imposes $\varphi \leq 1$, as one would expect of a parameter representing an efficiency. This result implies that the free energy released by the reaction is in part recycled in the system to increase the strain energy of the polymer chains, while the remainder is dissipated as heat.

We consider all species in the solvent to be at dilute concentration in water, *i.e.* $\Omega^k C^k \ll \Omega \, C$ for every species other than water, with $\Omega$ and $C$ the molar volume and molar concentration of water, respectively. The flux of solvent is then controlled by that of water, which carries all other species with it. Some migration of reactive species within the water flux might be considered as a consequence of chemical reactions. However, as will be explained later, within the time frame of consideration, there is no significant change in molar concentration of species, per unit solvent volume. Thus, we can consider the solvent as a homogeneous fluid and rewrite Eq. (4) as

$$J \approx 1 + \Omega \, C \tag{15}$$

Next, we describe the reaction kinetics as

$$q_c = r \, \Gamma^{\text{ATP}} \Gamma^m \Gamma^{\text{H}_2\text{O}} \tag{16a}$$

with

$$q_c = \frac{Q_c}{\Omega C} \tag{16b}$$

the reaction rate in moles of product per unit current solvent volume per unit time, and with

$$\Gamma^k \approx \frac{C^k}{\Omega C} \tag{16c}$$

the concentration of moles of species $k = \text{ATP}, m, \text{H}_2\text{O}$, per unit current solvent volume (where $m$ stands for "motors"). We can assume $\Gamma^{\text{H}_2\text{O}} \approx 1/\Omega$ given that water is the predominant species in the solvent. The reaction rate coefficient $r$ is obtained, as a function of temperature, via the Arrhenius relation $r = r_0 \exp(-E_A/RT)$, with $E_A$ the activation energy of the reaction (*i.e.* the energy barrier that must be crossed through thermal fluctuation to allow for the reaction to occur), and $r_0$ a constant that depends on the type of reaction. Substituting Eqs. (16b) and (16c) into (16a), we deduce that

$$Q_c \approx r\, \Gamma^{\text{ATP}} C^m / \Omega \tag{17}$$

where $C^m$ is the molar concentration of motors per unit reference volume, a quantity that is homogeneous and stationary. From experimental observations, given the constant velocity of the motors during active contraction [1], we assume the depletion of ATP molecules to be negligible, thus $\Gamma^{\text{ATP}}$ is constant with time. We also consider homogeneous distribution of ATP molecules within the solvent. If we also assume negligible temperature change within the gel, $r$ can be seen as homogeneous and stationary, and thus $Q_c$ is too. Finally, assuming the functional dependence $\varphi = \varphi(T, Q_c)$, we can also assume $\varphi$ to be constant. The rate of increase of cross-link density, from Eq. (14), now takes the form

$$\frac{dN}{dt} = p \Big/ \frac{\partial \psi}{\partial N} \tag{18a}$$

with

$$p \approx \varphi\, \Delta G^h\, r\, \Gamma^{\text{ATP}} C^m / \Omega \tag{18b}$$

the density of mechanical power, per unit reference volume, generated by the motors. As explained above, during gel contraction $p$ is approximately a constant.

The Helmholtz energy of the material can be modeled assuming additive contributions from Neo-Hookean elastic strain energy and enthalpy and entropy of mixing between solvent and polymer [8-9, 12-13],

$$\psi = \frac{1}{2} NkT(F_{ij}F_{ij} - 2\log J - 3) + \frac{RT}{\Omega}\left[\Omega C \log\left(\frac{\Omega C}{1+\Omega C}\right) - \frac{\chi}{1+\Omega C}\right] + \Pi(1 + \Omega C - J) \tag{19}$$

In this equation, $k$ is Boltzmann constant, $R$ is the gas constant, $\Pi$ is a Lagrange multiplier to enforce the molecular incompressibility of the solvent molecules and of the polymer chains. This also corresponds to the total pressure in the gel. Finally, $\chi$ is a constant describing the enthalpy of mixing.

Substitution of Eq. (19) into (7) gives the 1st Piola Kirchhoff stress as

$$t_{ij} = NkT\left(F_{ij} - F_{ji}^{-1}\right) - \Pi J\, F_{ji}^{-1} \tag{20}$$

while Eq. (19) into (8) gives the chemical potential of the solvent

$$\mu = RT\left[\ln\left(\frac{\Omega C}{1+\Omega C}\right) + \frac{\chi + 1 + \Omega C}{(1+\Omega C)^2}\right] + \Pi\, \Omega \tag{21}$$

The derivative of Eq. (19) with respect to *N* gives us

$$\frac{\partial \psi}{\partial N} = \frac{1}{2} kT \big( F_{ij} F_{ij} - 2 \log J - 3 \big) \tag{22}$$

and substitution of this in Eq. (18a) gives the rate of change of cross-link density, during gel contraction, as

$$\frac{dN}{dt} = \frac{2p}{kT \big( F_{ij} F_{ij} - 2 \ln J - 3 \big)} \tag{23}$$

Solvent diffusion can be described using Fick's law, in Lagrangian form [9, 12-13], as

$$J_i = -C \frac{D}{RT} F_{ik}^{-1} F_{jk}^{-1} \left( \frac{\partial \mu^{H_2O}}{\partial X_j} + \eta^{H_2O} \frac{\partial T}{\partial X_j} \right) \tag{24}$$

which, for any non-negative diffusivity coefficient, $D$, respects the condition imposed by Eq. (10). Boundary conditions for the mass diffusion problem are then,

$$\mu = \mu_{\text{ext}} \qquad \text{on } S_o \tag{25a}$$

if the boundary is permeable, where $\mu_{\text{ext}}$ is the chemical potential of the solvent molecules external to the gel and near the surface of it, or

$$J_i N_i = 0 \qquad \text{on } S_o \tag{25b}$$

if the boundary is impermeable. In a similar way, we can use Fourier's law to describe heat diffusion. However thermal effects are neglected in our treatment, assuming negligible temperature rise during the process. This assumption relies on the hypothesis of small specimen size, embedded in a large isothermal bulk solution, thereby allowing rapid removal of any heat generated within the gel. The temperature gradient in Eq. (24) is then zero and the temperature in Eq. (20) and (21) is uniform and constant.

We assume that gel contraction occurs without external loads except for the external pressure, $\Pi^{\text{ext}}$, of the solution within which the gel is embedded, and which must be equilibrated at the boundary in the deformed configuration. Equilibrium given by Eq. (1) then holds for

$$B_i = 0 \tag{26a}$$

$$T_i = -\Pi^{\text{ext}} J F_{ji}^{-1} N_j \tag{26b}$$

where we have neglected gravity. The initial conditions for contraction of the gel are associated with the undeformed state of the gel and thermodynamic equilibrium. The undeformed state of the gel corresponds to the condition of the swollen polymer with $F_{ij} = J^{1/3} \delta_{ij}$ everywhere in the body. This condition is associated with a homogeneous stress state that is in equilibrium, satisfying Eq. (1) along with Eq. (26), and thus

$$\Pi = \Pi^{\text{ext}} + NkT \big( J^{-1/3} - J^{-1} \big) \tag{27}$$

In this equation, the second term on the right hand side corresponds to the osmotic pressure that stretches the chains of the polymer network to accommodate the presence of the solvent.

The condition of thermodynamic equilibrium is defined by the absence of solvent flow. This, from Eq. (24), with constant and uniform $T$, and Eq. (25a), implies $\mu = \mu_{\text{ext}}$ everywhere in the gel. Once we calculate $\mu$ from Eq. (21), with substitution of $\Omega C$ from Eq. (15) and of $\Pi$ from Eq. (27), we obtain an algebraic equation in the variable $J$. The solution of this equation is $J_0$, the swelling ratio of the passive gel. We also consider $N_0$ to be the crosslink density of the passive gel. After activation, the gel stiffens and contracts based on the boundary conditions for solvent flow and gel deformation. Let us consider the gel to be free to deform in all directions and the external boundary of it to be permeable. The effect of the steady state contraction created by *static motors* can then be calculated as a new thermodynamic equilibrium state associated with a higher crosslink density $N_{ss} > N_0$. In this configuration we impose again $\mu = \mu_{\text{ext}}$ everywhere and obtain the swelling ratio $J_{ss} < J_0$ in the same way we obtained $J_0$. The shear modulus can be calculated in both states with Flory's formula [8] as $G = NkT\,J^{-1/3}$ by substitution of $J$ and $N$. Figure 2 reports the evolution of $G_{ss}$, the steady state shear modulus, as a function of $J_{ss}$ for various values of $J_0$ and $G_0$ (the initial shear modulus). In this figure, we reported $J_0$ of values 600, 800 and 1000, for $G_0 = 1\ Pa$ (black lines), and $G_0 = 10\ Pa$ (blue lines). The value of the constant $\chi$ is taken from *Appendix B*, where this case is analyzed for $J_0 = 1000$ and $G_0 = 1\ Pa$, observing a 10-fold increment in stiffness associated with a 30% deformation, as indicated in Figure 2.

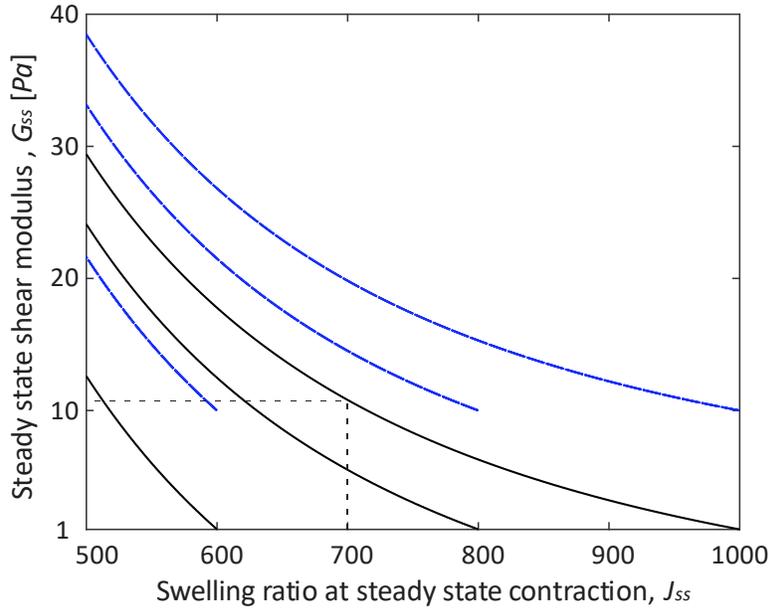

**Figure 2** Steady state swelling ratio $J_{ss}$ versus steady state shear modulus $G_{ss}$ for a gel contracted isotopically by static motors. We assumed initial swelling ratios $J_0$ of 600, 800, and 1000, and initial shear modulus $G_0$ of 1 *Pa* (black lines) and 10 *Pa* (blue lines).

During the activity of *dynamic motors*, we assume thermodynamic equilibrium is violated. The crosslink density is then $N > N_{ss}$, and it increases at the rate described by Eq. (23). This in turn increases the total pressure from Eq. (27), which creates an increment of the chemical potential in the body of the gel, given by Eq. (21), for which $\mu > \mu_{\text{ext}}$. The solvent molecules are progressively expelled from the gel, starting with the ones near the permeable boundary. The solvent flux determines the distribution of concentration of solvent via Eq. (5), but with the hydrolysis reaction giving negligible contribution, and thus $Q_c^{H_2O} \approx 0$.

Progressive stiffening terminates once all the dynamic motors detach from the shortened chains. We assume that the strain energy stored by a polymer chain can prompt this detachment, hence, the critical energy scales with the work required to detach a motor from its chain, $E^m$ (the energy of their intermolecular bonds). The energy stored by a chain scales with the energy stored in a portion of volume that includes one crosslink, $\psi^{st}/N$, with the strain energy in the polymer estimated as $\psi^{st} \approx \int_0^t (\partial \psi/\partial N)(\partial N/\partial t)dt = p\,t$ (neglecting the energy stored prior to contraction). We can then estimate the critical time when progressive stiffening stops from the relation

$$p\,t_c \sim N_c E^m \tag{28}$$

with $N_c$ the cross-link density at $t_c$, prior to the detachment of the dynamic motor. It can be observed from Eq. (18b) and (28) that a higher concentration of ATP will lead to a shorter contraction time $t_c$ with motors detaching sooner.

At this point the cross-link density goes back suddenly to its steady-state value $N_{ss}$ since the dynamic motors are no longer attached to the chains. Stresses, chemical potential and swelling ratio instead modify smoothly in time until thermodynamic equilibrium and steady-state conditions are restored. We consider for simplicity that each dynamic motor takes the same time interval to attach to a new chain and then repeats the process. In this way, the gel behavior is periodic with time and alternates between *contraction* and *recovery*, evidencing the peaks of gel displacement observed by Bertrand *et al.* [1].

*Uniaxial contraction*
We solve here the problem of uniaxial contraction of a slab of gel and use it as a model system for comparison with experiments [1]. These experiments consisted in measuring the movement of a bead sitting on top of a gel fragment. The inset of Figure 3 provides a sketch. We normalize the measured bead displacement by the steady state gel thickness $\ell$ (see *Appendix B*) to obtain the mean value of the nominal axial strain $e = \Delta \ell/\ell$. We represent the gel fragment as a slab, although the authors did not identify its shape [1]. The gel can deform only in direction $X_3$, which constrains the solvent flux to align with the vertical direction, hence $J_1 = J_2 = 0$. The slab sits on the surface of a glass plate and is in contact with a solution containing solvent molecules. The top of the gel constitutes a permeable free surface at $X_3 = 0$ where solvent molecules can diffuse freely. They diffuse from the gel to the solvent solution, in proximity of the top surface, and vice-versa, giving, from Eq. (25a)

$$\mu = \mu_{\text{ext}}, \qquad \text{at } X_3 = 0 \tag{29a}$$

The glass plate forms an impermeable boundary at $X_3 = L$, preventing any solvent flow at the bottom of the slab, giving, from Eqs. (25b) and (24), at constant $T$,

$$\frac{\partial \mu}{\partial X_3} = 0, \qquad \text{at } X_3 = L \tag{29b}$$

The axial stretch in the vertical direction is $\lambda_3$, while in the horizontal directions we have $\lambda_1 = \lambda_2 = \lambda_0$ at any time, with $\lambda_0 = J_0^{1/3}$. When the gel contracts we have

$$J = \lambda_0^2 \lambda_3 \tag{30}$$

From Eq. (20), the axial stress is then

$$t_{33} = NkT\,(\lambda_3 - 1/\lambda_3) - \Pi\,\lambda_0^2 \tag{31}$$

and is in equilibrium with Eqs. (1) and (26), from which we have $t_{33} = -\Pi^{\text{ext}}\,\lambda_0^2$ everywhere. This gives

$$\Pi = \Pi^{\text{ext}} + \frac{NkT}{J}(\lambda_3^2 - 1) \tag{32}$$

Substitution of Eqs. (15) and (32) into (21) gives

$$\frac{\mu - \Omega\,\Pi^{\text{ext}}}{RT} = \log\left(\frac{J-1}{J}\right) + \frac{\chi + J}{J^2} + \frac{n}{J}(\lambda_3^2 - 1) \tag{33a}$$

with

$$n = \frac{N\Omega}{N_a} \tag{33b}$$

and $N_a$ the Avogadro number. The chemical potential of the solvent molecules in the solution is $\mu_{\text{ext}} = \mu_0 + \Omega\,X(\Pi^{\text{ext}} - \Pi_0)$, where $\mu_0$ is the chemical potential of pure solvent at standard temperature and pressure, $X$ is the molar fraction of the solvent in the solution and $\Pi_0$ the standard pressure. Considering that water is the predominant species in the solution, we have $X \approx 1$, so that $\mu_{\text{ext}} - \Omega\,\Pi^{\text{ext}} \approx \mu_0 - \Omega\,\Pi_0$ which can be substituted in Eq. (33a) with $J = J_0$ to define the equilibrium state and give $\lambda_0 = J_0^{1/3}$.

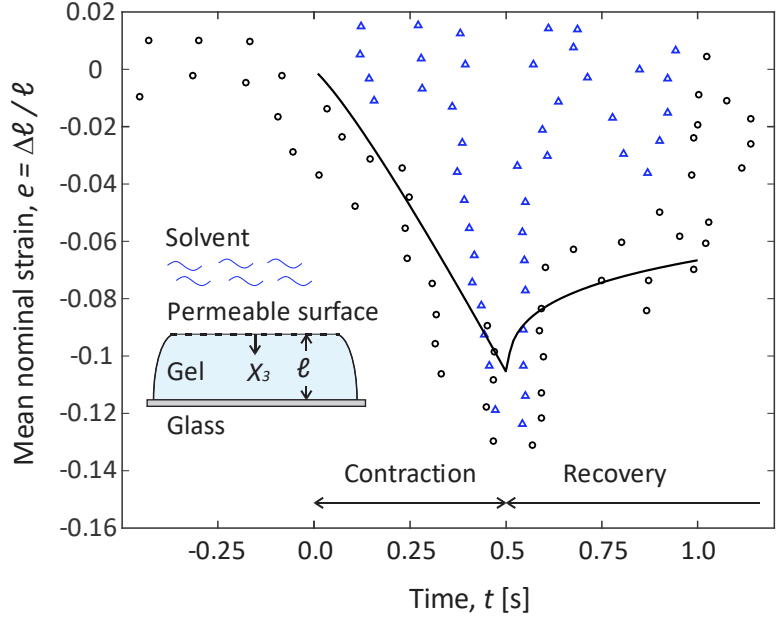

**Figure 3** Comparison of theoretical results (black line) with experiments from two distinct peaks in gel contraction (black circles and blue triangles) [1]. The inset sketches the system used to model uniaxial contraction for the theoretical results.

When the motors are activated, the static ones create a steady state chain shortening in the gel, making it stiffer and shorter. The shortening amount is about 30% [1], and therefore we consider $\lambda_{3,0} \approx 0.7\lambda_0$. Equating Eq. (30) with Eq. (15), differentiating with time and then substituting into Eq. (5) with $Q_c^{H_2O} \approx 0$ we have

$$\frac{d\lambda_3}{dt} \approx -\frac{\Omega}{\lambda_0^2}\frac{\partial J_3}{\partial X_3} \qquad (34)$$

Rewriting Eq. (24) for $J_3$ we have

$$J_3 = -\frac{D(\lambda_0^2\lambda_3-1)}{\Omega\lambda_3^2}\frac{\partial}{\partial X_3}\left(\frac{\mu}{RT}\right) \qquad (35)$$

Substituting Eq. (30) into (33a), then into (35) and into (34), we obtain

$$\frac{d\lambda_3}{dt} \approx D\frac{\partial}{\partial X_3}\left\{\left[\frac{1}{\lambda_0^4\lambda_3^4} - 2\chi\frac{\lambda_0^2\lambda_3-1}{\lambda_0^6\lambda_3^5} + n\frac{\lambda_0^2\lambda_3-1}{\lambda_0^4\lambda_3^4}(\lambda_3^2+1)\right]\frac{\partial \lambda_3}{\partial X_3} + \frac{\lambda_0^2\lambda_3-1}{\lambda_0^4\lambda_3^3}(\lambda_3^2-1)\frac{\partial n}{\partial X_3}\right\} \qquad (36)$$

Finally, Eq. (23) rewrites as

$$\frac{dN}{dt} = \frac{2p}{kT[2\lambda_0^2+\lambda_3^2-2\log(\lambda_0^2\lambda_3)-3]} \qquad (37)$$

Eq. (36) and (37) constitute a system of partial differential equations in the variables $\lambda_3$ and $N$. At $t = 0$, $\lambda_3 = \lambda_{3,0}$ and $N = N_{ss}$, which give the initial conditions for the system of equations. The boundary conditions can be defined by substitution of Eq. (30) into (33) and then into (29). The solution of this system of equations is obtained, using the physical parameters taken from literature [1, 9, 14-15] and described in *Appendix B*, to finally produce the results shown in Figure 3, 4 and 5.

Figure 3 compares the theoretical results with experiments [1], taken from two of the observed peaks of contraction. The experimental points (black circles and blue triangles) are scattered around a continuous trend, which agrees qualitatively with our simulation (black line). The scatter of the points is due to thermal fluctuation. Figure 4 reports the simulated distribution of the nominal axial strain through the thickness of the gel at different times during an extended (ideal) contraction time of 3.5 *s*. Figure 5 reports the simulated mean nominal axial strain *e*, as a function of time, for an extended contraction time of 5 *s* and for different values of power generation, from $0.1\,p$ to $10\,p$, with $p$ that adopted in Figure 3.

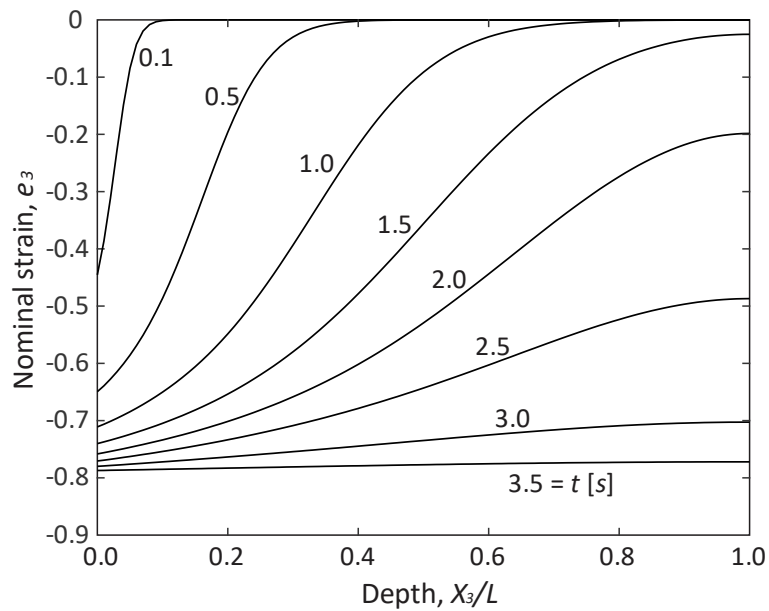

**Figure 4** Nominal strain distributions along the depth of the active gel obtained from the simulated results shown in Figure 3. The contraction time is here extended to an ideal value of 3.5 *s*.

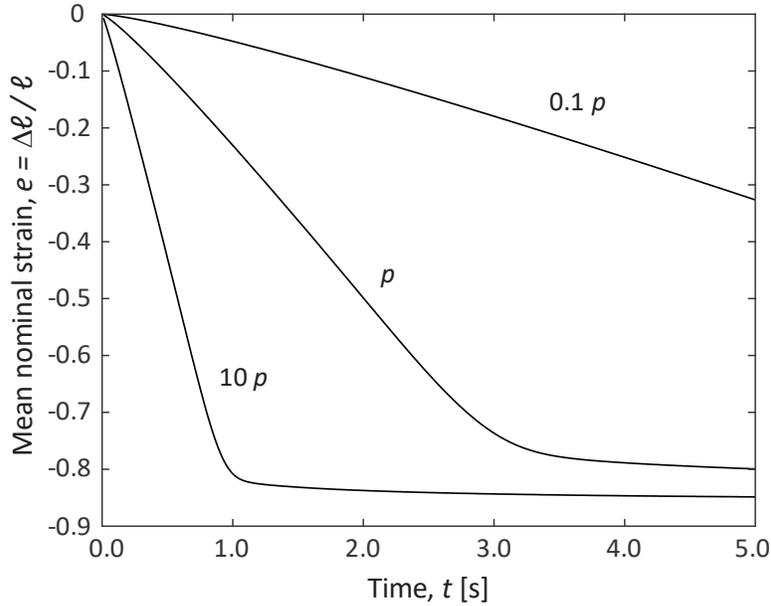

**Figure 5** Evolution of the mean nominal strain for various levels of mechanical power density generated by the molecular motors. The reference power density $p$ is adopted in the simulation results of Figure 3. The contraction time is here extended to an ideal value of 5 $s$.

## *Discussions and Conclusions*

As shown in Figure 3, our theory well predicts the trend in gel contraction and recovery. Deformation recovery (after the peak of contraction) is initially much faster than contraction, in agreement with the experiments [1, 3, 15], however it decelerates fast, leading to an overall slower recovery, compared to experiments. This might be due to the simplified model for describing the elastic behavior of the gel, based on neo-Hookean elasticity, which neglects enthalpic stiffness. Also, the experiments show a three-dimensional displacement of the bead, thus our hypothesis of uniaxial deformation generates an approximation. A more refined model, with use of finite elements, should be developed and implemented for a more realistic comparison. In Figure 4 we can observe that the contraction is initially localized in a region close to the free surface, where the solvent molecules can be expelled, and then progresses through the depth of the slab. At 3.5 $s$, the contraction becomes homogeneous in the gel as it progresses. Figure 5 shows an initial stage with a high rate of mean strain of contraction (line slope), corresponding to the initial stage at Figure 4. A second stage with much slower contraction starts when the strain becomes homogenous along the depth (Figures 3-4). If molecular motors were to be designed to maximize contraction, their power generation and contraction time should be maximized. The latter parameter would find its optimum at the point of transition between the first and the second regimes, given the second regime produces little contraction. This transition, though, depends on the geometry of the specimen and on the diffusivity of the solvent through the gel and its interface with the surroundings.

The ratio of static motors to dynamic motors is thus far postulated in the model due to the lack of experimental evidences. The effect of static motors is used to formulate the initial conditions for the boundary value problem. The hypothesis of simultaneous detachment and attachment of all dynamic motors represent another simplification in our model. The number of attached dynamic motors per unit volume is the variable $C^m$ in Eq. (17) and (18b) and its evolution with time and its distribution in space affects the evolution of the power density $p$ in Eq. (18a). Albeit simultaneous attachment of motors is probably unrealistic, simultaneous detachment results intuitive from the following reasoning. Take a small control volume in the gel. The average strain energy stored by the polymer chains in that volume increases with time by motor activity at the rate defined by the parameter $p$. This process also increments the osmotic pressure in the gel, which is associated with the elastic resistance of the polymer network to swelling. When the critical conditions for the detachment of the first motor are met, the chain at which it was attached will suddenly release all its tension and local movement of solvent molecule will follow in order to redistribute chain tension in the control volume. This will increment the tension in the chains that still have motors attached, hence prompting further motor detachment in cascade. The timescale for the described mechanism is proportional to that of local (short-range) motion of solvent molecules, which we consider to be negligibly small compared to the timescale of all other phenomena described. In order to improve the model by accounting for non-simultaneous attachment of motors, one should develop a statistical law for the evolution of $C^m$ that accounts for the attachment probability of a given motor, at a given time.

*Acknowledgement*

The work at UBC has been supported by the Natural Science and Engineering Research Council of Canada (NSERC), under the Discovery Grant. The Institute for Computing, Information and Cognitive Systems (ICICS), at the University of British Columbia, provided logistic support. Work at UCSB was supported by the U.S. Department of Energy (DOE), Office of Science, Basic Energy Sciences (BES) under Award DESC0014427.

[*]mbacca@mech.ubc.ca

***Appendix A***

The first law of thermodynamics is given by

$$\frac{d}{dt}\int_{V_o} e\, dV_o = -\int_{S_o} N_i J_i^h\, dS_o + \int_{S_o} T_i v_i\, dS_o + \int_{V_o} B_i v_i\, dV_o - \int_{S_o} N_i\, h^k J_i^k\, dS_o \tag{A1}$$

where $e$ is internal energy per unit volume in the reference state, $J_i^h$ is heat flux in the reference configuration, $v_i$ is the velocity of elements of the polymer network and thus is the rate of change of $x_i$ and $h^k$ is the partial molar enthalpy of species $k$. Use of the divergence theorem and the principle of virtual power gives us

$$\frac{de}{dt} = -\frac{\partial J_i^h}{\partial X_i} + t_{ij}\frac{dF_{ij}}{dt} - \frac{\partial\left(h^k J_i^k\right)}{\partial X_i} \tag{A2}$$

The rate of change of entropy of the body is

$$\frac{d}{dt}\int_{V_o} \eta\, dV_o = -\int_{S_o} N_i \frac{J_i^h}{T}\, dS_o - \int_{S_o} N_i\, \eta^k J_i^k\, dS_o + \int_{V_o} \dot{\eta}^p\, dV_o \tag{A3}$$

where $\eta$ is the entropy per unit volume in the reference state, $\eta^k$ is the partial molar entropy of species $k$ and $\dot{\eta}^p$ is the rate of entropy production per unit volume in the reference state. This leads to

$$\frac{d\eta}{dt} = -\frac{1}{T}\frac{\partial J_i^h}{\partial X_i} - J_i^h \frac{\partial}{\partial X_i}\left(\frac{1}{T}\right) - \frac{\partial\left(\eta^k J_i^k\right)}{\partial X_i} + \dot{\eta}^p \tag{A4}$$

The Helmholtz energy per unit volume is given by

$$\psi = e - T\eta \tag{A5}$$

and thus

$$\frac{d\psi}{dt} = t_{ij}\frac{dF_{ij}}{dt} - \frac{\partial(h^k J_i^k)}{\partial X_i} - \frac{J_i^h}{T}\frac{\partial T}{\partial X_i} + T\frac{\partial(\eta^k J_i^k)}{\partial X_i} - \eta\frac{dT}{dt} - T\dot\eta^p \quad (A6)$$

We introduce $\mu^k$, the chemical potential of species $k$, and observe that

$$\mu^k = h^k - T\eta^k \quad (A7)$$

The rate of change of the specific Helmholtz energy, by us equating Eq. (A5) to the time derivative of Eq. (A5) and by us using Eq. (5), becomes

$$\frac{d\psi}{dt} = t_{ij}\frac{dF_{ij}}{dt} - \left(\frac{\partial\mu^k}{\partial X_i} + \eta^k\frac{\partial T}{\partial X_i}\right)J_i^k - \frac{J_i^h}{T}\frac{\partial T}{\partial X_i} + \mu^k\left(\frac{dC^k}{dt} - Q_c^k\right) - \eta\frac{dT}{dt} - T\dot\eta^p \quad (A8)$$

We next assume that the Helmholtz energy has functional dependence provided by Eq. (6) and rewrite Eq. (A8) as an expression for entropy production

$$T\dot\eta^p = \left(t_{ij} - \frac{\partial\psi}{\partial F_{ij}}\right)\frac{dF_{ij}}{dt} + \left(\mu^k - \frac{\partial\psi}{\partial C^k}\right)\frac{dC^k}{dt} - \left(\eta + \frac{\partial\psi}{\partial T}\right)\frac{dT}{dt} - \left(\frac{\partial\mu^k}{\partial X_i} + \eta^k\frac{\partial T}{\partial X_i}\right)J_i^k - \frac{J_i^h}{T}\frac{\partial T}{\partial X_i} - \mu^k Q_c^k - \frac{\partial\psi}{\partial N}\frac{dN}{dt}$$
(A9)

From the second law of thermodynamics, we must impose the right hand side of Eq. (A9) to be $\geq 0$.

Considering chemical equilibrium, homogenous and stationary distribution of temperature and species concentration, which also implies no flux of species, and no change of cross-link density, the first 3 terms in parenthesis in Eq. (A9) must be $\geq 0$ for all deformation rates, concentration changes and temperature adjustments, whether positive or negative. This gives Eqs. (7), (8) and (9). Eq. (A9) can then be rewritten as

$$T\dot\eta^p = -\left(\frac{\partial\mu^k}{\partial X_i} + \eta^k\frac{\partial T}{\partial X_i}\right)J_i^k - \frac{J_i^h}{T}\frac{\partial T}{\partial X_i} - \mu^k Q_c^k - \frac{\partial\psi}{\partial N}\frac{dN}{dt} \quad (A10)$$

If we assume that the first, the second and the third and fourth together are independent processes, from Eq. (A10) we then obtain inequalities at Eqs. (10), (11), and (12).

### *Appendix B*

Considering room temperature, we have $kT = 4.14 \cdot 10^{-21} J$ and $RT = 2.49 \cdot 10^3 J$. The linear dimension of a water molecule is 3 Å, giving a solvent molar volume of $\Omega \approx 1.63 \cdot 10^{-5} m^3$. The diffusion coefficient is estimated using the Stokes-Einstein formula $D = kT/(6\pi\rho\eta)$, where $\rho$ is the radius of a water molecule, 1.5 Å, and $\eta = 8.9 \cdot 10^{-4} Pa\ s$ is the viscosity of water, giving finally $D = 1.65 \cdot 10^{-9} m^2/s$.

Bertrand *et al.* [1] measured the shear modulus of the passive gel as $G_0 = 1\ Pa$. Adopting the relation $G_0 = kT/R_g^3$, we can estimate the radius of gyration of the linkers, giving $R_g \approx 160\ nm$. The Kuhn length of one linker chain is reported as $L_L \approx 2000\ nm$, while the cross sectional area is $A_L \approx 1\ nm^2$ [1]. This gives a volume of polymer, per crosslink, estimated as $V_s \approx 3A_L L_L = 6 \cdot 10^3\ nm^3$. The total volume of gel, per crosslink, is $V \approx R_g^3 = 4 \cdot 10^6\ nm^3$, giving a volume fraction of polymer of $f_s \approx$ $10^{-3}$. From this we estimate the initial swelling ratio as $J_0 = 1/f_s \approx 10^3$, giving $\lambda_0 = J_0^{1/3} \approx 10$ for the passive gel. The shear modulus of the polymer network before swelling is $N_0 kT$, while that of a gel subject to uniaxial contraction is $G = NkT\lambda_3/\lambda_0^2$. For the passive gel we have $\lambda_3 = \lambda_0$ and $N = N_0$, giving $G_0 = N_0 kT/\lambda_0$ [8]. From this we obtain the crosslink density of the passive gel as $N_0 \approx 2.42 \cdot 10^3 \, \mu m^{-3}$.

After activation of the motors, Bertrand *et al.* [1] observed a 10-fold increase in gel stiffness at steady state, *i.e.* the stiffening created by static motors, giving $G_{ss} = 10 \, G_0$. This was also accompanied by a contraction of $\sim 30\%$, giving $\lambda_{3,0} = 0.7 \, \lambda_0$, and then $N_{ss} = G_{ss}\lambda_0^2/kT\lambda_{3,0} = 3.46 \cdot 10^4 \, \mu m^{-3}$. At steady state, we consider the chemical potential of the solvent to be homogenous and equal to that of the passive gel. Thus, Eq. (33) must give the same result for $J = \lambda_0^3$ and $N = N_0$, and for $J = \lambda_{3,0}\lambda_0^2$ and $N = N_{ss}$. This condition is satisfied only if we choose $\chi = 0.44$. The value of initial and steady state chemical potential of the solvent is then calculated, from Eq. (33), as $\mu_{ext} - \Omega \, \Pi^{ext} = \mu_0 - \Omega \, \Pi_0 \approx 5.4 \cdot 10^{-8} RT = -1.34 \cdot 10^{-4} \, J/mol$.

The size of the gel fragments was not measured in Bertrand *et al.* [1], however, they observed it to be in the range of $1 \, \mu m$ and estimated the relation $A/z \sim 3 \, \mu m$, with $A$ and $z$ the area and the thickness of the fragment, respectively. We adopt $z = 0.6 \, \mu m$ and $A = 1.8 \, \mu m^2$ so that the volume of the fragment is $\sim 1 \, \mu m^3$. $z$ constitutes the initial length of the fragment, giving reference length $L = z/\lambda_0 = 0.06 \, \mu m$, and a characteristic time $t^* = L^2/D = 2.2 \cdot 10^{-4} s$. The thickness of the gel fragment at steady state is $\ell = 0.7z = 0.42 \, \mu m$ and the bead displacements observed in the experiments [1] are divided by this length to obtain the experimental values of the mean nominal strain $e$ reported in Figure 3.

A single motor FtsK50C has been shown to produce forces up to $50 \, pN$ and to travel along the chain at a speed of more than $1.7 \, \mu m/s$ [1,14-15], which gives a power generation, per motor, of $p^m \approx 10^{-4} pW$. Let us consider a value of $p = 1.4 \, pW/\mu m^3$, to observe the same maximum contraction as in the experiments [1], so that we use this as a calibration parameter. Considering $p = p^m C^m$, we have $C^m \approx 1.4 \cdot 10^4 \mu m^{-3}$, which corresponds to a mean motor spacing of $L^m \approx 0.04 \, \mu m$ in the reference state, and $l^m \approx 0.4 \, \mu m$ in the swollen state (giving roughly 16 dynamic motors per gel fragment). The contraction time $t_c$ can be estimated using Eq. (28). Considering the motor-chain intermolecular bonds being primarily hydrogen-type, we estimate $E^m \approx 10^{-6} \, pJ$. Given $N_0 \approx 2 \cdot 10^3 \mu m^{-3}$, and making the assumption of 700-fold maximum stiffening, as obtained in our numerical results, we have $N_c \approx 10^6 \mu m^{-3}$. Finally we obtain a value in the order of $t_c \approx 0.5 \, s$ which also confirms the experimental observations in Bertrand *et al.* [1], where the average contraction time is $0.5 \, s$ (in Figure 3 it can be observed that the peak corresponding to the black circles has slightly longer contraction time while in that corresponding to the blue triangles is shorter).